\documentclass[letterpaper,10pt]{article} 

\usepackage{opticameet3} 

\usepackage[locale=US]{siunitx}
\usepackage{threeparttable, nicefrac}

\newcommand\authormark[1]{\textsuperscript{#1}}

\usepackage{amsmath,amssymb, booktabs}
\usepackage[colorlinks=true,allcolors=black]{hyperref}

\begin{document}

\title{3D printed human skull phantoms for transcranial photoacoustic imaging}

\author{Hannah Linde\authormark{ 1}, Saskia Menzer\authormark{ 1}, Jan Laufer\authormark{ 1}, Thomas Kirchner\authormark{ 1}\authormark{*}}

\address{\authormark{1} Institut für Physik, Martin-Luther-Universität Halle-Wittenberg, Halle (Saale), Germany}

\email{\authormark{*} thomas.kirchner@physik.uni-halle.de} 

\begin{abstract*}
Photoacoustic (PA) waves are strongly distorted and attenuated in skull bone. To study these effects on PA imaging, we designed and 3D-printed tissue-mimicking phantoms of human skull. We present a comparison of results in phantom and \emph{ex vivo} skull.
\end{abstract*}

\section{Introduction} \label{sec:introduction}
Photoacoustic (PA) brain imaging has the potential to be translated into a real-time non-ionizing clinical modality which could provide  faster diagnosis and monitoring for time-critical brain injuries, compared to slower current clinical imaging modalities like X-ray computer tomography (CT) and magnetic resonance imaging (MRI).\cite{Yao.2014}
Applications of PA measurements of the human brain through the intact skin and skull could include monitoring of brain hemodynamics, the detection of intracranial bleeding, and supporting stroke diagnosis \cite{Yao.2014, Wang.2023}.

The strong acoustic attenuation and wavefront distortion introduced by cranial bone tissue pose significant challenges for biomedical photoacoustic imaging \cite{Huang.2012, Liang.2021} as they result in reduced image resolution and contrast \cite{Na.2020}.
In previous work \cite{Leonov.2022}, tissue-mimicking human bone phantoms were fabricated using a combination of fused deposition modelling (FDM) and stereolithography (SLA), producing SLA resin-printed temporal bone phantoms for PAI. That work suggested that materials used in FDM printing are not sufficiently transmissive to ultrasound. High acoustic attenuation is actually desirable in skull phantoms as ultrasound above approximately $2\,\textrm{MHz}$ does not penetrate thick human skull\cite{Kirchner.2023}. A disadvantage of SLA 3D printing is its inability to create hollow structures, such as the diplo\"e pores in thick skull bone, without drainage holes \cite{Xu.2024}. Another approach to mimicking skull in experiments is the use of a spherical acrylic shell as a skull phantom \cite{Poudel.2020}. Acrylic is an elastic solid that possesses acoustic properties similar to bone but it lacks the micro-structure of the diplo\"e and therefore does match the acoustic aberration of thick skull.

In this work, we developed a skull-mimicking phantom using FDM 3D printing, which allows us to fabricate a water-permeable, porous medium.
As FDM 3D printing allows for rapid prototyping, we iteratively optimized printing parameters, resulting in a printed phantom of a human frontal cranial bone which has acoustic properties closely mimicking tissue.

\section{Materials and methods} \label{sec:material_methods}
We created a printable mesh model of a representative right frontal cranial bone based on a bone segmentation obtained from an X-ray micro-CT data set\cite{Kirchner.2022}. The corresponding skull bone of a 70-year-old male body donor was used as the reference for our printed frontal bone phantoms. Thick skull bone, like the frontal cranial bone (general thickness of 6 to \SI{9}{\milli\meter}), is composed of an outer and an inner layer of solid cortical bone separated by a cancelous bone layer -- the diplo\"e -- which accounts for \SI{54(15)}{\percent} of the skull thickness \cite{Alexander.2019}.

\begin{figure}[bht]
    \centering
    \includegraphics{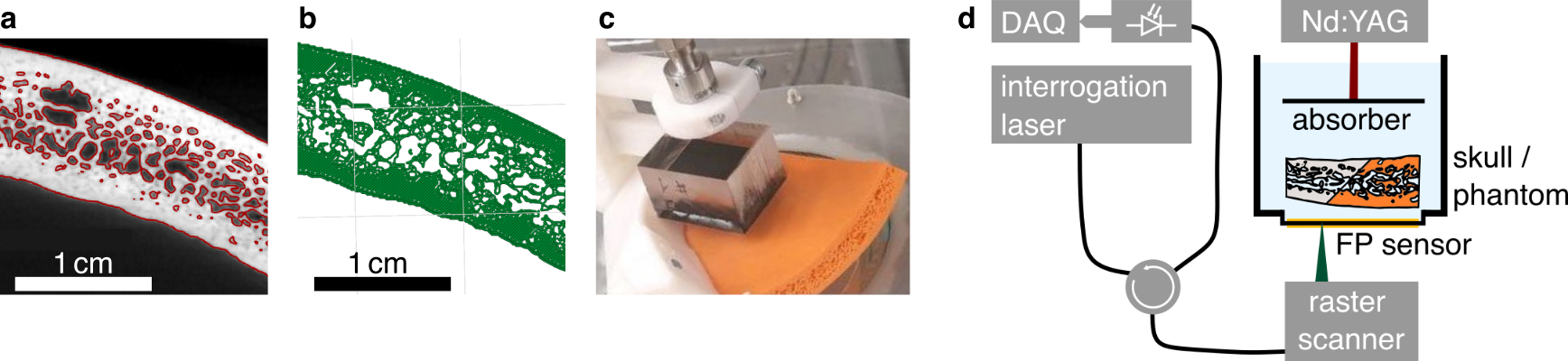}
    \caption{\textbf{a} example slice of the micro-CT with bone segmentation boundary in red, \textbf{b} print path for the same slice, \textbf{c} resulting printed phantom (orange) during characterization. \textbf{d} Transmission measurement setup. PA imaging through \emph{ex vivo} human skull and skull phantoms. The planar PA source, a black-coated acrylic glass block, is illuminated by an Nd:YAG excitation laser and generates PA wavefronts. These propagate through a region of frontal cranial bone or phantom and are measured by a planar Fabry-Perot (FP) sensor, raster-scanned by an interrogation laser.}\label{fig:skullseg}
\end{figure}

We generated the mesh from the bone segmentation using the Medical Imaging Interaction Toolkit (MITK)\cite{Wolf.2005}. We limited the mesh size due to computational constraints by reducing the mesh vertices to \SI{25}{\percent} without any relevant loss in detail. Example sections of a micro-CT slice and the corresponding bone segmentation are shown in \autoref{fig:skullseg}a. A section of one slice of the resulting print path is shown in \autoref{fig:skullseg}b. 15 phantom versions were printed based on this mesh (see \hyperref[sec:dataavailability]{Data availability}) with varying printing parameters. Acoustic transmission measurements and PA imaging were performed through the resulting 3D-printed phantoms and the corresponding \emph{ex vivo} skull.

\subsection*{3D printing of skull phantoms} \label{subsec:print_skull_lookalike}
The phantoms were printed with Polylactide (PLA) filament, the most widely used 3D printing filament. We printed the phantoms with varying resolution (printing nozzle size and layer height) and porosity (by artificially enlarging the diplo\"e pores). For the latter we used the \emph{hole compensation} method in the slicing software. To mimic solid cortical bone, the infill density was generally set to \SI{100}{\percent}.
Full print settings can be found in the \hyperref[sec:dataavailability]{Data availibility} section. The phantom print paths were prepared using a slicing software (Bambu Studio, Bambu Lab, Shenzhen, China) and printed using a widely used printer (Bambu X1 Carbon, Bambu Lab, Shenzhen, China).

The phantoms were printed with varying formulations of PLA: PLA Basic, PLA Matte or PLA Tough (Bambu Lab, Shenzhen, China). \autoref{tab:material_properties} compares the mechanical properties of PLA (as provided by the manufacturer) to cranial bone. The speed of sound was measured by ultrasound transmission through massive printed PLA slabs.

\begin{table}[h]
    \centering
    \caption{Mechanical properties of human skull bone \cite{Auperrin.2014, Roberts.2013, Pichardo.2011} compared to PLA filament}
    \label{tab:material_properties}
    \begin{tabular}{l|cc}
        \toprule
             & Cranial bone & PLA \\
        \midrule
            Density [$\nicefrac{g}{cm^3}$] & 1.71 $\pm$ 0.13 & 1.27 $\pm$ 0.03 \\
            Elastic modulus [GPa] & 3.81 $\pm$ 1.55 &  2.38 $\pm$ 0.36 \\
            Tensile strength [MPa] & 67.73 $\pm$ 17.8 & 33.67 $\pm$ 7.36\\
            Bending modulus [GPa] & 11.73 $\pm$ 0.95 & 2.68 $\pm$ 0.35\\
            Bending strength [MPa] & 82.0 $\pm$ 25.5 & 72 $\pm$ 9.99\\
            Speed of sound [m/s] & 2504 $\pm$ 120 & 2115 $\pm$ 4\\
        \bottomrule
    \end{tabular}
\end{table}

\subsection*{Photoacoustic transmission measurements and imaging} \label{subsec:measuring_phantoms}
As illustrated in \autoref{fig:skullseg}d, a layer of black acrylic paint on a PMMA substrate is used as a PA source. The emitted PA wavefronts are measured after passing through either \emph{ex vivo} human skull or the printed phantoms.

We used a Fabry-Perot (FP) raster scanning tomograph setup for single point measurements and PA imaging. The scanner is based on Zhang \emph{et al.} \cite{Zhang.2008} and incorporates a broadband planar FP sensor as described by Buchmann \emph{et al.} \cite{Buchmann.2017}. We compare the acoustic power spectra of the insertion losses in the samples (skull phantoms, \emph{ex vivo} human skull). A Nd:YAG excitation laser (Quanta-Ray PRO-270-50H, Spectra-Physics Lasers, Santa Clara, USA) was used for PA excitation at \SI{1064}{\nano\meter} with pulse durations of $\sim$\SI{10}{\nano\second} and a pulse repetition rate of \SI{50}{\hertz}. Tomographic data sets were acquired by raster scanning the interrogation beam across a detection aperture of $\SI{1.6}{\centi\meter} \times \SI{1.6}{\centi\meter}$ in \SI{0.1}{\milli\meter} steps using galvanometer mirrors (GVS012, Thorlabs) and recording the time-resolved PA signals at each position.

To avoid air inclusions, the water tank was filled with degassed and deionized water in which the samples were submerged for at least 15 minutes. The PA source was positioned on the same acoustic axis as the sensor at a distance of $\sim$\SI{35}{\milli\metre}. The source was illuminated by the collimated output of the excitation laser via a multimode fiber. The pulse energy was attenuated to $\sim$\SI{2}{\milli\joule} and the spot size diameter measured $\sim$\SI{2}{\milli\meter}.

We acquired PA image data sets for all samples in water (and without the samples as reference).
The PA image volumes were reconstructed with the k-Wave toolbox \cite{Treeby.2010} using a fast Fourier reconstruction algorithm\cite{Kostli.2001} assuming a speed of sound of \SI{1480}{\metre\per\second}. 

\section{Results and discussion} \label{sec:results_discussion}

\begin{figure}[h]
    \includegraphics{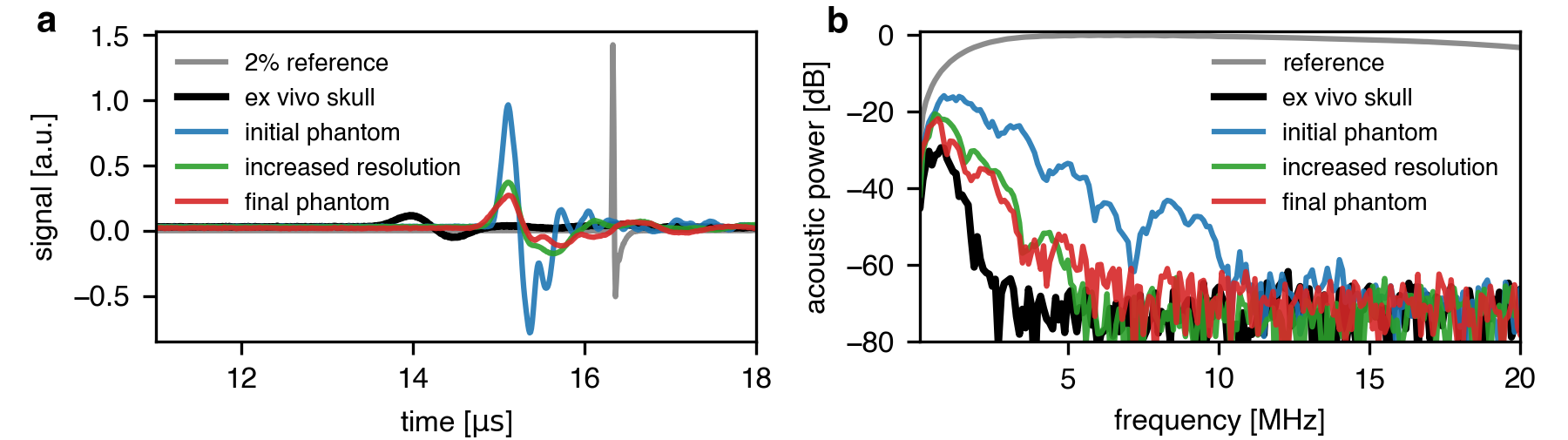}
    \caption{\textbf{a} Representative PA waveforms measured in a single point of an FP sensor \textbf{b} Corresponding acoustic power spectra -- discrete Fourier transforms of (a).}
    \label{fig:main_results}   
\end{figure}

The printing parameters most effective in optimizing acoustic attenuation are the decrease in nozzle diameter to 0.2\,mm, using a 0.1\,mm layer height and the increase in porosity by using a 0.1\,mm hole compensation -- the parameters for the phantom marked as \emph{final}. The measurements for the phantom can be seen in \autoref{fig:main_results}, including the reference measurement through water, \emph{ex vivo} human skull and two further phantoms.

\begin{figure*}[b!]
    \centering
    \includegraphics{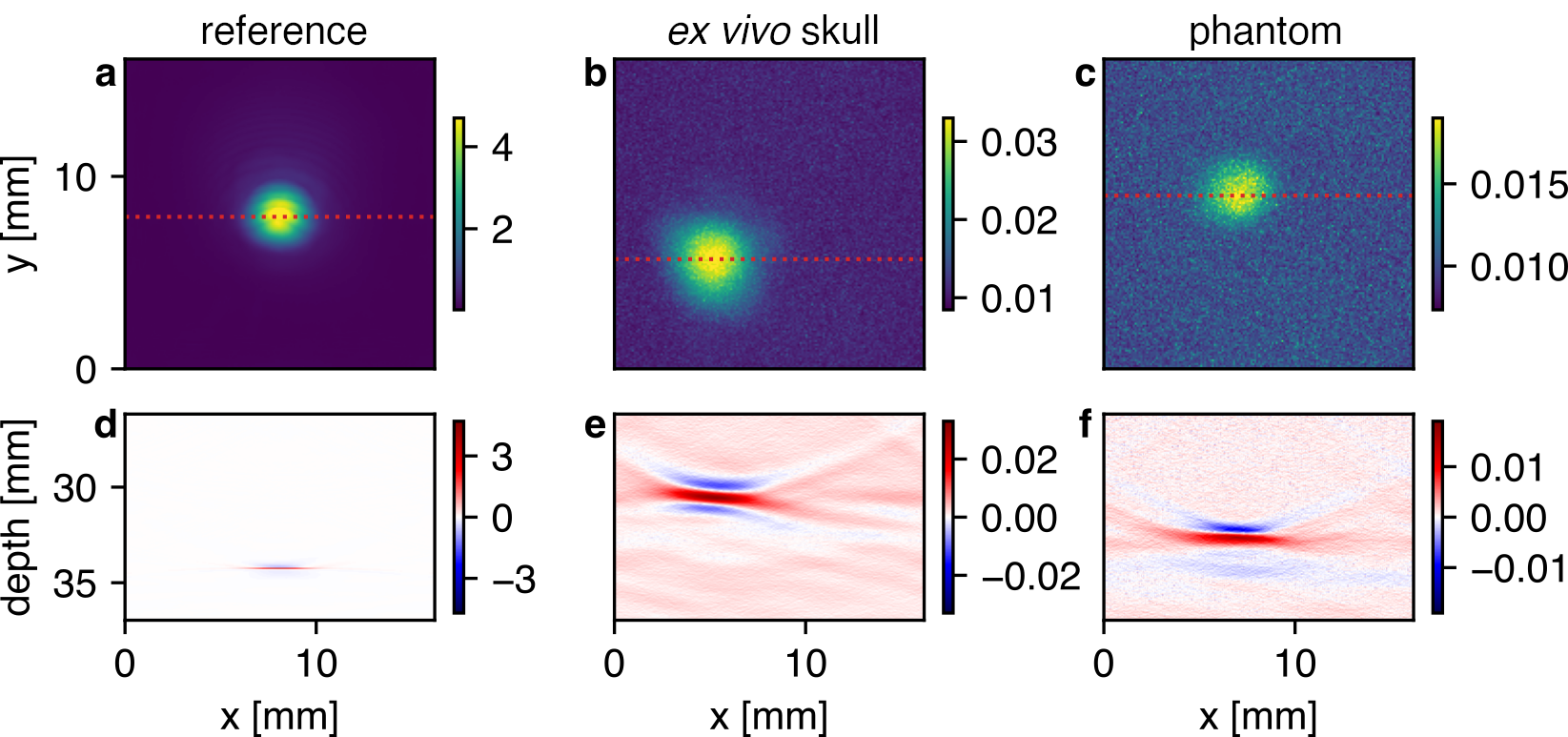}
    \caption{PA imaging of a collimated laser spot projected onto a layer of black acrylic paint. (\textbf{a\&d}) shows the reconstructed PA image through water with no skull, (\textbf{b\&e}) through \emph{ex vivo} frontal cranial bone and (\textbf{c\&f}) through the \emph{final} skull phantom. (\textbf{a}-\textbf{c}) maximum intensity projections (MIP) along the depth axis of the reconstructed PA images. (\textbf{d}-\textbf{f}) Selected longitudinal slices at the highlighted position (red dotted line in figures above). All figures are shown in a linear color scale.}
    \label{fig:MIP}
\end{figure*}

The measurements on \emph{ex vivo} skull show large attenuation of signal components above $\sim$\SI{2}{\mega\hertz}. The \emph{initial phantom} did not match this attenuation, likely because the printing resolution (\SI{0.4}{\milli\meter}) was too low to accurately recreate the microstructure of the skull. Therefore, walls separating porous structures were not printed correctly, as these walls in the diplo\"e are around \SIrange{0.1}{0.3}{\milli\meter} thick \cite{Alexander.2019}. Increasing the printing resolution to \SI{0.2}{\milli\meter} greatly improved the phantom's performance (\emph{increased resolution} phantom). 
By increasing the porosity of the phantom artificially using the \emph{x-y-hole compensation} the attenuation of the phantom was increased further. The \emph{final} phantom printed with both increased resolution and improved porosity shows a reasonable match in acoustic attenuation compared to the \emph{ex vivo} skull sample. Increasing the hole compensation even further led to merging pores and lower effective attenuation. Scaling up the phantom size (i.e. by \SI{20}{\percent}), did not significantly change the acoustic attenuation. In initial tests, various infill patterns were evaluated, with no significant effect. The presented phantoms all use a rectilinear infill pattern and PLA Matte as the printing material.

Using the PA tomography setup, we also imaged the PA source through the \emph{final} phantom and the \emph{ex vivo} human skull. In \autoref{fig:MIP} maximum intensity projections (MIP) of an example image volume are shown (a-c) as well as example slices (d-f). A qualitative comparison of PA imaging through skull versus phantom shows a similar distortion and attenuation. 

In conclusion: We presented a 3D-printed phantom which mimics \emph{ex vivo} skull in both single point insertion loss measurements as well as in PAI. The phantom is easy to modify, adapt and fabricate as it is Open Data and made with widely available 3D printers and materials.

\section*{Data availability}\label{sec:dataavailability}
    PAI volume data is available at \href{https://doi.org/10.5281/zenodo.14876160}{doi:10.5281/zenodo.14876160}, together with the 3D printable mesh files and parameters. The X-ray Micro-CT data set is Kirchner (2022) \cite{Kirchner.2022}. Data processing and analysis was primarily performed using the open-source k-Wave toolbox as described.

\section*{Acknowledgements}
    The authors would like to thank Heike Kielstein, Institute of Anatomy and Cell Biology, MLU Halle-Wittenberg for the provision of the human skull sample. 
    This work was funded by the German Research Foundation (DFG, Deutsche Forschungsgemeinschaft) under grant number 471755457.

\section*{Credit authorship contribution statement}
    Hannah Linde:
Data curation,
Formal analysis,
Investigation,
Methodology,
Validation,
Visualization,
Writing – original draft,
Writing – review \& editing. 
Saskia Menzer:
Formal analysis,
Investigation,
Validation,
Writing – review \& editing. 
    Jan Laufer: Methodology, Resources,  Writing – review \& editing.
    Thomas Kirchner: 
Conceptualization
Funding acquisition, 
Methodology, 
Project administration, 
Software, 
Resources, 
Supervision, 
Validation, 
Visualization, 
Writing – original draft,
Writing – review \& editing.

\bibliographystyle{elsarticle-num}
\bibliography{Literatur.bib}

\end{document}